\documentclass[11pt,a4paper]{article}
 \pdfoutput=1
\makeatletter
\newcommand*{\centernot}{%
  \mathpalette\@centernot
}
\def\@centernot#1#2{%
  \mathrel{%
    \rlap{%
      \settowidth\dimen@{$\m@th#1{#2}$}%
      \kern.5\dimen@
      \settowidth\dimen@{$\m@th#1=$}%
      \kern-.5\dimen@
      $\m@th#1\not$%
    }%
    {#2}%
  }%
}
\date{}
\makeatother
\newcommand{\independent}{\perp\mkern-8.mu\perp}
\newcommand{\notindependent}{\centernot{\independent}}

\usepackage[latin1]{inputenc}
\usepackage{amsmath}
\usepackage{amssymb}
\usepackage{amsfonts}
\usepackage{graphicx}
\usepackage{color}
\usepackage{soul}
\usepackage{mathrsfs}  
\usepackage{url}  
\usepackage{natbib}

\usepackage{color}

\date{}

\begin{document}

\def\spacingset#1{\renewcommand{\baselinestretch}%
{#1}\small\normalsize} \spacingset{1}


  \title{\bf Adjusting for bias introduced by instrumental variable estimation in the Cox Proportional Hazards Model}

 \author{Pablo Mart\'inez-Camblor\thanks{{\it Correspondence to:} Pablo Mart\'inez-Camblor. TDI, Dartmouth College, 7 Lebanon Street, Suite 309, Hinman Box 7251, Hanover, NH 03755, 
USA. E-mail: {\color{blue} Pablo.Martinez.Camblor@Dartmouth.edu}}
\hspace{.2cm}\\
 \small{The Dartmouth Institute for Health Policy and Clinical Practice,}\\  \small{Geisel School of Medicine at Dartmouth, Hanover, NH, USA}\\
\and
  Todd A. MacKenzie\\
\small{The Dartmouth Institute for Health Policy and Clinical Practice,}\\  \small{Department of Biomedical Data Science},\\  \small{Geisel School of Medicine at Dartmouth, Hanover, NH, USA}\\
\and
     Douglas O. Staiger\\
\small{The Dartmouth Institute for Health Policy and Clinical Practice,}\\  
\small{Department of Economics, Dartmouth College, Hanover, NH, USA}\\
\and
Philip P. Goodney\\
\small{The Dartmouth Institute for Health Policy and Clinical Practice,}\\ 
\small{ Department of Vascular Surgery,  Dartmouth Hitchcock Medical Center}\\
\small{Geisel School of Medicine at Dartmouth, Hanover, NH, USA}\\
\and
  A. James O'Malley\\
\small{The Dartmouth Institute for Health Policy and Clinical Practice,}\\  \small{Department of Biomedical Data Science},\\  \small{Geisel School of Medicine at Dartmouth, Hanover, NH, USA}\\
}  \maketitle
\begin{abstract}
Instrumental variable (IV) methods are widely used for estimating average treatment effects in the presence of unmeasured confounders. However, the capability of existing IV procedures, and most notably the two-stage residual inclusion (2SRI) procedure recommended for use in nonlinear contexts, to account for unmeasured confounders in the Cox proportional hazard model is unclear. We show that instrumenting an endogenous treatment induces an unmeasured covariate, referred to as an individual frailty in survival analysis parlance, which if not accounted for leads to bias. We propose a new procedure that augments 2SRI with an individual frailty and prove that it is consistent under certain conditions. The finite sample-size behavior is studied across a broad set of conditions via Monte Carlo simulations. Finally, the proposed methodology is used to estimate the average effect of carotid endarterectomy versus carotid artery stenting on the mortality of patients suffering from carotid artery disease. Results suggest that the 2SRI-frailty estimator generally reduces the bias of both point and interval estimators compared to traditional 2SRI.\\
\noindent%
{\it Keywords:}   Two-stage residual inclusion; Individual frailty effect; Unmeasured confounding.  
\end{abstract}
\vfill

\spacingset{1.45} 

Ever since Cox's seminal paper in 1972 \citep{Cox72}, the Cox proportional hazards models has become one of the most widely used statistical models due to the ubiquity in medicine of time-to-event outcomes subject to censoring. Although 
the Cox model was traditionally applied to analyze small data sets from randomized clinical trials (RCT), the increasing cost of RCTs and increasing availability of observational data has led to increased utilization of the Cox model in non-randomized settings. A clever observational study, perhaps in conjunction with a small RCT, can overcome the necessity for a large and highly expensive RCT. Registries containing the procedures and outcomes of all patients with a particular diagnosis (e.g., the Surveillance, Epidemiology, and End Results (SEER) program of the National Cancer Institute, \url{https://seer.cancer.gov/about/}), or that have undergone a certain procedure (e.g., the Vascular Quality Initiative, VQI, registry, \url{http://www.vascularqualityinitiative.org/}), are typically designed to measure all known risk factors and so yield high quality observational data. Nonetheless, unmeasured confounding is a concern whenever randomization is lacking. Instrumental variable (IV) methods \citep{angrist96} allow causal interpretations from an analysis of observation data. They may also be applied to RCTs with imperfect compliance in which the objective is to estimate the effect of treatment received (the average treatment effect of the treated) rather than just intention-to-treat.  However, even if a valid IV is available (e.g. treatment assigment in a RCT), an unresolved question in statistics and econometrics is the best way of using an IV with the Cox model. In this paper we seek to answer this question.\par

There are many real-world applications in which hazard ratio (HR) estimates from RCTcomparisons and Cox models estimated on observational data differ even after controlling for detailed clinical information on patients, suggesting the need to develop IV procedures for Cox models. For example, large differences were observed in the HR of death for carotid endarterectomy versus carotid artery stenting estimated using the Cox model adjusted for covariates in the VQI and the results of recent randomized clinical trials \citep{nejm, otru}. The VQI analysis finds an estimated hazard ratio (HR) of 0.693 (95\% confidence interval: (0.633; 0.760)), implying longer survival times under endarterectomy, whereas both clinical trial found almost null effects (approximate HR of 0.97, 95\% CI (0.93; 1.02) and 0.91 (0.69; 1.20), respectively). The profound difference in these results may be due unmeasured confounders that have a strong effect on the procedure a VQI patient receives and their subsequent survival time. Therefore, the development of an IV procedure for survival time data that yields RCT-like estimates for a more general VQI population is highly desired.\par

\begin{figure}[b!]
\begin{center}
\includegraphics[width=8cm]{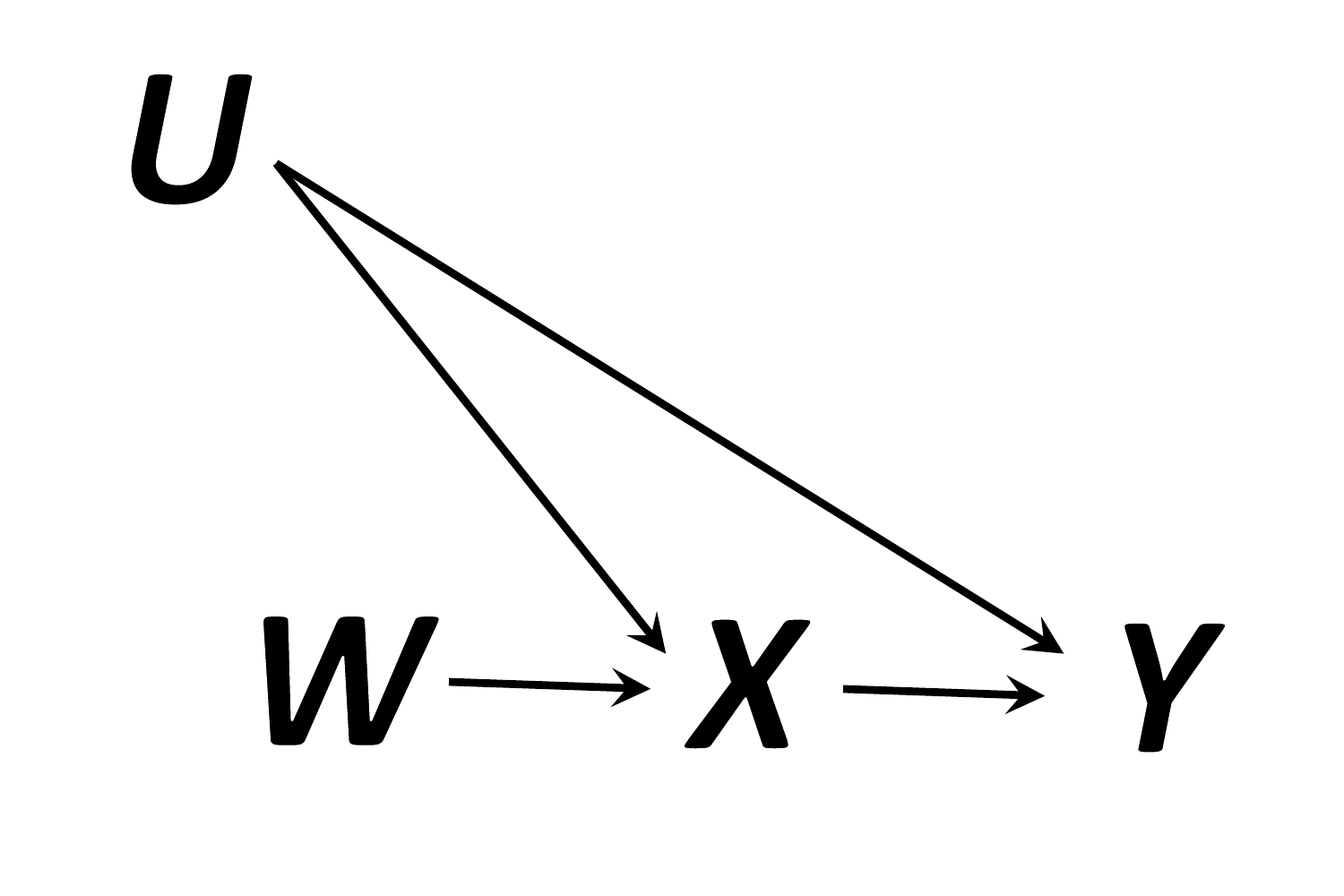}
\end{center}
\caption{DAG with unmeasured confounder ${\boldsymbol U}$, 
treatment ${\boldsymbol X}$, and outcome $Y$. The IV ${\boldsymbol W}$ is related with
$Y$ only through ${\boldsymbol X}$.}
\label{F1}
\end{figure}

In the context of structural linear equations, the two-stage least squares estimator, 2SLS, was introduced in econometrics in the 1950s 
by various authors (see \cite{R1} and references therein). IV procedures have subsequently been used to estimate causal effects from observational studies \citep{R2} in statistics, biomedicine, and many other applied research fields. The Directed Acyclic Graph (DAG) \citep{R3} in Figure \ref{F1} presents the basic IV identification strategy of conditioning on the IV ${\boldsymbol W}$ instead of the treatment ${\boldsymbol X}$ in order to block the flow of association from the unmeasured confounder ${\boldsymbol U}$ to the outcome ${\boldsymbol Y}$ through ${\boldsymbol X}$, thereby identifying the causal effect of ${\boldsymbol X}$ on the outcome $Y$.\par

Conventional instrumental variable methods can produce substantial estimation bias when the true underlying models are nonlinear \citep{R4}. 
The direct nonlinear generalization to 2SLS is the two-stage predictor substitution, 2SPS, procedure \citep{R6}. In the first stage, the relationship between 
the IV, ${\boldsymbol W}$, and the exposure, ${\boldsymbol X}$, is estimated by any consistent estimation technique. Then, the resulting fitted exposure status replaces
the real observed exposure in the outcome model.\par

Alternatively, \cite{R7} proposed the two-stage residual inclusion, 2SRI, or  {\it control function estimator}. The 2SRI computes the expected exposure as for 2SPS but for the second-stage 
augments the target model with the residuals from the first-stage. The first and second stage models can be linear or nonlinear. Although
there exists some debate about the relative performance of the 2SPS and 2SRI procedures, 2SRI is generally considered to have theoretical and practical advantages over 2SPS
\citep{R8, R9}. \cite{cai} compared the bias of the 2SPS and 2SRI procedures at estimating the odds ratio among compliers under the principal stratification 
framework. They found that 2SRI is asymptotically unbiased in the absence of unmeasured confounding but bias increasingly occurs with the magnitude of unmeasured confounding 
while 2SPS was always biased. See \cite{R5} for an extensive review of different IV procedures for both linear and nonlinear contexts.\par

The estimation of treatment effects by IVs for time-to-event outcomes has received attention recently. The challenge is the presence of right censoring and non-linearity.
Due to the presence of additive effects and an explicit error term, accelerated failure time models \citep{R10} and additive hazard models (see, for instance, \cite{R11}, \cite{R12} and references therein) 
are the most common frameworks. However, in biomedical and epidemiological research proportional hazard models are overwhelmingly used to analyze time-to-event outcomes.\par

\begin{figure}[b!]
\begin{center}
\includegraphics[width=8cm]{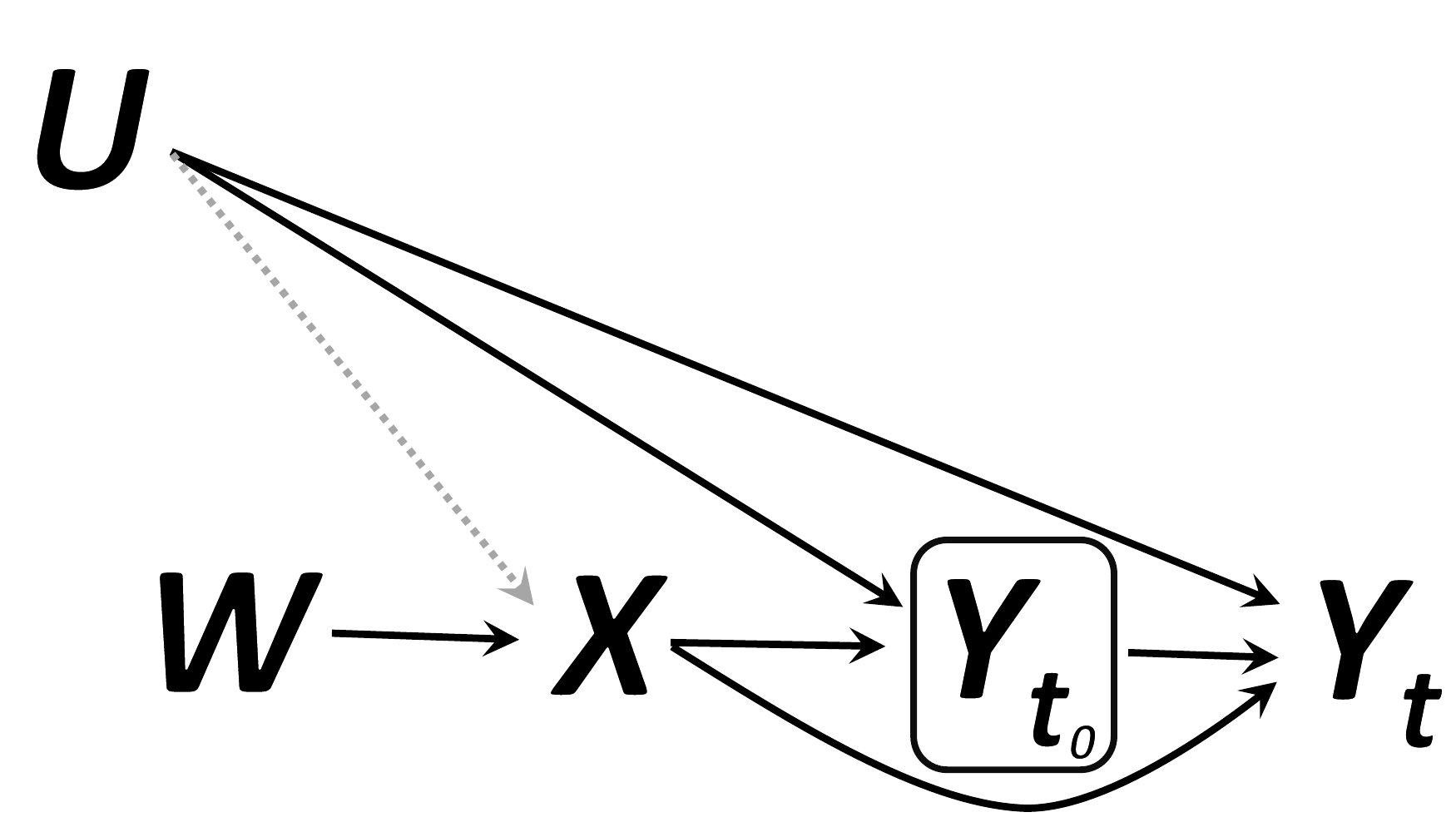}
\end{center}
\caption{Directed acyclic graph, DAG, showing the unmeasured confounder ${\boldsymbol U}$, treatment ${\boldsymbol X}$, and the time-to-event outcome $Y$ at $t_0$ and $t=t_0 + \epsilon$ where $\epsilon$ represents an arbitrarily small amount of time. 
The independent variable ${\boldsymbol W}$, related with $Y_{t_0}$ and $Y_t$ only through ${\boldsymbol X}$ is an instrumental variable.}
\label{F2}
\end{figure}

Most practitioners are familiar with Cox's proportional hazard model \citep{Cox72} and interpreting research 
results in terms of hazard ratio. An IV estimator of the hazard ratio has been proposed that assumes that the omitted covariates have an additive effect on the hazard \citep{todd1}. A consistent estimator of the hazard ratio has been derived for the setting of a binary instrument (e.g. randomization), \citep{todd2}. However, from a causal standpoint estimating the hazard ratio is problematic \citep{R13}. Because the hazard function is the instantaneous change in the survival probability, survival status $Y_{t_{0}}$ just prior the instantaneous time period is conditioned on thereby inducing an association between treatment and any unmeasured predictor of survival (Figure~\ref{F2}), a phenomena known as collider bias.  Intuitively, the problem is that even if ${\boldsymbol X}$ is independent of ${\boldsymbol U}$ at time 0, it is not independent at any time $t>0$ because selection out of the sample depends on ${\boldsymbol U}$.

A consequence of collider bias is that if the path ${\boldsymbol U}\rightarrow {\boldsymbol X}$ exists then blocking it using an IV is not sufficient to identifying the causal estimate for the HR of ${\boldsymbol X}$ on $Y_t$, bringing the performance of IV methodology at estimating HRs into question. This is seen from evaluations of the bias of both the 2SPS 
and the 2SRI procedures on causal hazard ratio estimation under Weibull models \citep{cai} and in other contexts \citep{wan}, which found that no procedure consistently performed the best and that the bias of the IV estimator of the causal hazard ratio depended on the form of the unmeasured confounder, and magnitude of its effect, and the data structure.\par

Even if ${\boldsymbol U}$ does not have a causal effect on ${\boldsymbol X}$, it is known that under the proportional hazard assumption the true effect of measured covariates on the hazard-ratio is underestimated if ${\boldsymbol U}$ is ignored \citep{R26, R27}. Therefore, in survival time models, omitted predictors unrelated with treatment assignment should be accounted for to avoid misspecification \citep{R15}, a contrast to a linear regression model where an explicit error term would absorb ${\boldsymbol U}$. Because omitted predictors of survival that are unrelated with any of the measured predictors manifest as frailties, the addition of an individual frailty \citep{R17} appears to be a possible remedy.\par

The crux of the research in this paper rests on the observation that if (i) a continuously valued exposure, $X$, is related to an instrumental variable $W$ and a covariate $U$ by the linear model, $X=\alpha_W\cdot W + \alpha_U\cdot U + \epsilon$, and (ii) the effects of $X$ and $U$ on a time-to-event, $Y$, satisfies the Cox model, ${\cal  P}\{Y \geq t|X=x, U=u\}=\exp\{-\Lambda_0(t)\exp\{\beta_X\cdot x + \beta_U\cdot u\}\}$, then the conditional distribution of the time-to-event given the exposure, $X$, and  $R=X-\alpha_W\cdot W$ satisfies the Cox model with frailty term: ${\cal P}\{Y \geq t|X=x, R=r\}={\cal P}\{Y \geq t|X=x, \alpha_U\cdot U + \epsilon=r\}={\cal P}\{Y \geq t|X=x,  U = \alpha_U^{-1} \cdot (r - \epsilon)\}={\mathbb E}_{\epsilon}[{\cal P}\{Y \geq t|X=x,  U = \alpha_U^{-1}\cdot  (r - \epsilon)\}]={\mathbb E}_{\epsilon}[\exp\{-\Lambda(t)\cdot {\boldsymbol z}\cdot \exp\{\beta_X\cdot x + \beta_U\cdot \alpha_U^{-1}\cdot r\}\}]$ where ${\boldsymbol z}=\exp\{-\beta_U\cdot \alpha_U^{-1}\cdot \epsilon\}$ is the frailty term. The above expectation has the form of a Cox model with a frailty, suggesting that 2SRI will work in the context of Cox's model if the second stage is implemented with a frailty. Specifically, the procedure requires that the distribution specified for the frailty matches the distribution of the noise term in the linear model of $X$ given $W$ and $U$. This strategy assumes that $\alpha_U\neq 0$, the omitted covariates are related with treatment assignment, and that $\alpha_W$ is known thereby motivating evaluation of what happens when $\alpha_W$ (and hence ${\boldsymbol X}$) is estimated, when ${\boldsymbol X}$ is binary, and when the wrong frailty distribution is assumed.\par

The novel contributions of this paper are to: (i) confirm the above reasoning that 2SRI applied to the Cox model induces a frailty, (ii) derive a 2SRI algorithm with an individual frailty term in the second stage estimation, (iii) explore the performance of the procedure both when ${\boldsymbol X}$ is binary (as opposed to continuous), and (iv) evaluate the robustness of the procedure to misspecification of the frailty distribution. The remainder of the paper is structured as follows. The notation and assumed underlying model are defined in Section 2. Theoretical justification of the 2SRI-frailty procedure is provided 
in Section 3. Section 4 evaluates the operating characteristics of the new 2SRI-frailty procedure using Monte Carlo simulations. We consider the common situation in which we have 
a binary treatment with continuous IV and continuous measured and unmeasured covariates. We first consider the case where the exact same unmeasured predictors are present in both the treatment selection and the survival time models. We then consider the more realistic scenario where there are related but different unmeasured confounders in the survival and the treatment selection models, in which the first scenario is a special case. In Section 5, the proposed methodology is applied to an observational study where the goal is to estimate the average effect on mortality of the treatment (carotid endarterectomy (CEA) versus carotid artery stenting (CAS)) received by patients suffering from carotid artery disease. The paper concludes in Section 6. 

\section{Notation and models}
\label{sec:notation}

Conventionally, the right-censored framework assumes the sample $\{(t_i,\delta_i)\}_{i=1}^N$, where $t_i=\min\{y_i,\, c_i\}$ and $\delta_i=I_{(-\infty,c_i]}(y_i)$ ($I_A(x)$ stands for the usual indicator 
function, taking value $1$ if $x\in A$ and $0$ otherwise), and
$c_i$ and $y_i$ are the censoring and the survival times for the $i$th subject ($1\leq i\leq N$), respectively. Let $S_Y(\cdot)={\cal P}\{Y>\cdot\}$ denote the survival function of $Y$. The Cox proportional hazard model is given by
\begin{equation}
d[\log S_Y(t|X,{\boldsymbol Z},{\boldsymbol U})]=-\lambda_0(t)\cdot\exp\{\beta_0+\beta_X\cdot X +{\boldsymbol\beta^t_Z}\cdot {\boldsymbol Z}+{\boldsymbol\beta^t_{\boldsymbol U}}\cdot {\boldsymbol U}\},
\end{equation}
where $\lambda_0(\cdot)$ is the baseline hazard function, $X$ is a random variable representing the study treatment, ${\boldsymbol Z}$ is a random vector of exogenous measured predictors and ${\boldsymbol U}$ is a random vector of unmeasured predictors. 

The goal is to estimate the value of $\beta_X$; that is, the average change in the risk of an individual caused by a change in the received treatment. \par
\medskip
We assume that an individual's received treatment is the result of the selective process depicted by the equation,
\begin{equation}
X=\alpha_0 + {\boldsymbol\alpha^t_W}\cdot {\boldsymbol W} + {\boldsymbol\alpha^t_Z}\cdot {\boldsymbol Z}+{\boldsymbol\alpha^t_V}\cdot {\boldsymbol V} + \epsilon,
\end{equation} 
\label{eq1}
with ${\boldsymbol W}$ a measured random vector and ${\boldsymbol V}$ a random vector of 
unmeasured variables that may be correlated with unmeasured variables in ${\boldsymbol U}$. The term $\epsilon$ is an 
independent random component representing the difference between the variable $X$ and the predicted values obtained in the linear model. Note that the uncontrolled relationship between the unmeasured covariates 
affecting survival and treatment selection operate through the relationship between ${\boldsymbol U}$ 
and ${\boldsymbol V}$ and/or ${\boldsymbol Z}$. If ${\boldsymbol V}\independent{\boldsymbol U}$, the survival model just 
contains unmeasured covariates, ${\boldsymbol U}$, that are not unmeasured confounders.

If the random vector ${\boldsymbol W}$ satisfies:
\begin{enumerate}
\item[$C_1$.] ${\boldsymbol W}\notindependent (X|\boldsymbol{ Z,\, U,\, V})$,
\item[$C_2$.] ${\boldsymbol W}\independent (Y|X,\,\boldsymbol{Z,\, U})$ (exclusion restriction assumption),
\item[$C_3$.] ${\boldsymbol W}\independent\boldsymbol{(V,\, U|Z)}$ (randomization assumption),
\end{enumerate} 
then ${\boldsymbol W}$ can be considered an {\it instrumental variable}. The strength of the instrument is reflects the strength of the relationship between ${\boldsymbol W}$ and $X$. In a randomized trial with perfect compliance, assigned treatment is a perfect instrument. Assumptions $C_1$, $C_2$ and $C_3$ can be reformulated and combined with the 
stable unit treatment value assumption, SUTVA, and the monotonicity assumption \citep{R18} between the treatment to complete the conditions under which the IV ${\boldsymbol W}$ identifies the causal effect of ${\boldsymbol X}$ non-parametrically (i.e., without relying on (1) and (2)). Common instruments include prior institutional affinity for using a particular procedure, geographic region of residence, an individual's differential access to certain treatments, and an individuals genes (aka Mendelian randomization, \cite{R19}).

\section{Proposed methodology}
\label{sec:method}
The proposed methodology considers a standard first stage in which the relationship among the treatment, $X$, the 
instrument variable, ${\boldsymbol W}$, and the measured confounding, ${\boldsymbol Z}$, is estimated by any 
consistent method. We use simple linear regression models with standard least squares estimation but more flexible procedures could also be implemented. The first stage procedure is:
\begin{enumerate}
\item[Step 1.] From the data, estimate the parameters to obtain the predicted values
 $$\hat X=\hat\alpha_0 + \hat {\boldsymbol\alpha}_{\boldsymbol W}^t\cdot {\boldsymbol W} + \hat {\boldsymbol\alpha}_{\boldsymbol Z}^t\cdot {\boldsymbol Z}.$$
 Then, compute the residuals, $\hat R=X -\hat X$.
\end{enumerate}

It is worth noting that, under model (\ref{eq1}) and assuming ${\boldsymbol Z}\independent\boldsymbol{V}$, 
$\hat R$ is a consistent estimator of $[{\boldsymbol\alpha^t_V}\cdot {\boldsymbol V} +\epsilon ]$. If 
${\boldsymbol Z}\notindependent\boldsymbol{U}$,  $\hat R$  estimates 
$[({\boldsymbol\alpha_Z}-\hat {\boldsymbol\alpha}^t_{\boldsymbol Z})\cdot {\boldsymbol Z} + {\boldsymbol\alpha^t_V}\cdot {\boldsymbol V} +\epsilon ]$. 
Almost sure convergence of $\hat {\boldsymbol \alpha}_{\boldsymbol Z}$ to ${\boldsymbol \alpha_Z}$ is not guaranteed. The residual $\hat R$ contains all the information about the unmeasured vector ${\boldsymbol U}$ related with the treatment assignment and 
unrelated with ${\boldsymbol Z}$; that is, all the available information about unmeasured confounding is contained in the residuals provided by the model (1). However, $\hat R$ also contains {\it white noise} pertaining to idiosyncratic or purely random factors affecting an individual's treatment selection, which corresponds to the difference between the 
unmeasured covariates in the two models, ${\boldsymbol V}$ and ${\boldsymbol U}$, and to the 
independent random term $\epsilon$ in model (2). We conjecture that the component of $\hat R$ due to white noise 
can be handled by specifying an individual {\it frailty} in the outcome model in order to allow $\hat R$ to more fully be able to perform its intended task of controlling for ${\boldsymbol U}$. The proposed second stage is:
\begin{enumerate}
\item[Step 2.] Estimate the Cox proportional hazard regression with individual frailty:  
$$d[\log\hat S_Y(t|X,{\boldsymbol Z},\hat R,F)]=-z\cdot\hat\lambda_0(t)\cdot\exp\{\hat\beta^*_0+\hat\beta^{\text {IV}}_X\cdot X +\hat {\boldsymbol\beta}^{*\,t}_{\boldsymbol Z}\cdot {\boldsymbol Z}+\hat\beta_{\hat R}\cdot\hat R\},$$
where $z=\exp\{F\}$ is the individual frailty term. A distribution should be specified for $F$ (e.g., log-Gaussian, Gamma). The parameter estimate of $\beta_X$ that results from this procedure is denoted $\hat\beta^{\text {IV}}_X$.
\end{enumerate}

Standard algorithms for estimating Cox models with frailties may be used to implement the procedure. For example, \cite{R21} proved that maximum likelihood estimation for the Cox model with a Gamma frailty can be accomplished using a general penalized routine, and \cite{RN1} derived a similar argument for the Cox model with a Gaussian frailty. 

\subsection{Asymptotic properties of $\hat\beta^{\text {IV}}_X$}
We derive the asymptotic distribution of the 2SRI-frailty estimator 
$\hat\beta^{\text {IV}}_X$ for the case in which 
${\boldsymbol U}={\boldsymbol V}$. If ${\boldsymbol U}\neq {\boldsymbol V}$ a similar derivation can be performed by decomposing 
${\boldsymbol U}$ and ${\boldsymbol V}$ into common and orthogonal terms and making standard reliability assumptions on the distributions of these terms.\par
\medskip
By adapting the convergence results for Cox's partial likelihood (see, for instance, Theorem 5.3.1 in \cite{R20}) we obtain the following convergence results.\par
\medskip\noindent
{\bf Theorem.} Assume the causal models
\begin{align}
\label{t1}
d[\log S_Y(t|X,Z,U)]&=-\lambda_0(t)\cdot\exp\{\beta_0+\beta_X\cdot X + \beta_Z\cdot Z+\beta_U\cdot U\},\\ 
\label{t2}
X&=\alpha_0 + \alpha_W\cdot W + \alpha_Z\cdot Z+\alpha_U\cdot U + \epsilon,
\end{align}
with the random variables $X$ (the treatment), $Z$ (measured covariate) and $U$ (unmeasured covariate). In addition, assume that $U$ is normally distributed, 
that $\epsilon$ is independently normally distributed random noise, and that $W$ (the instrument) is a random variable satisfying $C_1$-$C_3$. Then, if the 
censoring time, $C$, satisfies $C\independent (Y| X,\, Z,\, U,\, W)$, we have the weak convergence,
\begin{equation}
\sqrt{n}\cdot\left\{\hat\beta^{\text {IV}}_X - \beta_X \right\}\stackrel{\mathcal L}{\longrightarrow}\,\mathscr N\left(0,\sigma^{\text {IV}}_X\right),
\end{equation}
where $\left(\sigma^{\text {IV}}_X\right)^2=n\cdot\mathbb V[\hat\beta^{\text {IV}}_X]$ ($\mathbb V$ stands for the variance operator) can be consistently estimated 
from the survival and  at-risk counting processes.\par
\medskip\noindent
{\it Proof.} In absence of the frailty term, it is well-known that the estimator of $\boldsymbol\beta$ that maximizes the Cox-model partial likelihood function obeys the asymptotic law
\begin{equation}
\sqrt{n}\cdot\{\hat{\boldsymbol\beta}-{\boldsymbol\beta}\}\stackrel{\mathcal L}{\longrightarrow}\,\mathscr N_p\left(0,{\boldsymbol I^{-1}({\boldsymbol \beta})}\right),
\end{equation}
where ${\boldsymbol\beta}=\{\beta^*_0,\beta_X,\beta^*_Z\}$, and ${\boldsymbol I({\boldsymbol \beta})}$ is the $p\times p$ ($p$ stands for dimension of the vector 
${\boldsymbol\beta}$) information matrix \cite{R20}, which can be consistently estimated
by ${\boldsymbol I(\hat {\boldsymbol \beta})}$. 

In the presence of an individual frailty, different estimation methods have been proposed. Particularly, \cite{RN1} and \cite{RN2} studied the case of multiplicative log-normal distributed frailties. The proposed methodology obtains maximun-likelihood 
estimates of the regression parameters, the variance components and the baseline hazard, as well as empirical Bayes estimates of the 
random effects (frailties). Therefore, it suffices to prove that the second stage of our proposed 2SRI-frailty procedure is a Cox proportional hazard model with a gaussian 
frailty term in which the coefficient related with the treatment is unchanged from the original model (i.e., $\beta_X$).

 Given the causal equation in (\ref{t2}) and $C_1-C_3$, $S_1$ yields
$$\hat R= (\alpha_0-\hat\alpha_0) + (\alpha_Z-\hat\alpha_Z)\cdot Z + \alpha_U\cdot U + \epsilon + {\cal O}(n^{-1/2}).$$
Hence, $\alpha_U\cdot U=\hat R -\{ (\alpha_0-\hat\alpha_0)+(\alpha_Z-\hat\alpha_Z)\cdot Z + \epsilon + {\cal O}(n^{-1/2})\}$. If $\alpha_U=0$, the 
linear part of the model in (\ref{t1}) can be rewritten as,
\begin{align*}
{\mathfrak L}=&\beta_0+\beta_X\cdot X +\beta_Z\cdot Z+\beta_U\cdot U\\
             =&\beta^*_0+\beta_X\cdot X +\beta^*_Z\cdot Z+\hat R + F,
\end{align*}
where $\beta^*_0=\beta_0 - (\alpha_0-\hat\alpha_0)+\beta_U\cdot\mathbb E[U]$, $\beta^*_Z=\beta_Z- (\alpha_Z-\hat\alpha_Z)$ and $F=\beta_U\cdot (U-\mathbb E[U]) + \epsilon + {\cal O}(n^{-1/2})$. 
Due to $\epsilon$ and $U$ being independent normally distributed variables, $F$ is also asymptotically independent and normally distributed. If $\alpha_U\neq 0$, then 
\begin{align*}
{\mathfrak L}=\beta^*_0+\beta_X\cdot X +\beta^*_Z\cdot Z+\beta_{\hat R}\cdot\hat R + F,
\end{align*}
with $\beta_{\hat R}=\beta_U/\alpha_U$, and where, in this case, $\beta^*_0=\beta_0 - \beta_{\hat R}\cdot (\alpha_0-\hat\alpha_0)$, 
$\beta^*_Z=\beta_Z- \beta_{\hat R}\cdot (\alpha_Z-\hat\alpha_Z)$ and $F=\beta_{\hat R}\cdot \epsilon + {\cal O}(n^{-1/2})$. Therefore, $F$ is asymptotically 
 independent and normally distributed due to $\epsilon$ being independent and normally distributed. Hence, the survival model is given by
$$d[\log S_Y(t|X,Z,U)]=-\lambda_0(t)\cdot z\cdot\exp\{\beta^*_0+\beta_X\cdot X +\beta^*_Z\cdot Z+\beta_{\hat R}\cdot\hat R\},$$ 
\noindent
which has the form of a Cox proportional hazards model with frailty $z=\exp\{F\}$. Therefore, if $\hat {\boldsymbol\beta} =\{\hat\beta^*_0,\hat\beta^{\text IV}_X,\hat\beta^*_Z\}$ is the estimator 
resulting from step $S_2$, invoking the censoring time assumptions and using the convergence of the partial maximum-likelihood method given a consistent method of estimating the product of the baseline risk and the frailty \citep{RN1, RN2}, it follows that
\begin{equation}
\sqrt{n}\cdot\left\{\hat\beta^{\text {IV}}_X - \beta_X \right\}\stackrel{\mathcal L}{\longrightarrow}_n\,\mathscr N\left(0,\sigma^{\text {IV}}_X\right),
\label{result}
\end{equation}
with $(\sigma^{\text {IV}}_X)^2$ the component in the matrix ${\boldsymbol I^{-1}({\boldsymbol \beta})}$ corresponding to $\beta_X$.\hfill{$\square$}\par
\medskip
{\noindent\bf Remark.} Normality of $U$ is required only when $\alpha_U=0$. In this case, the survival model 
does not contain unmeasured confounders, just unmeasured covariates. Such {\it white noise} can be omitted in standard 
linear models, but not in Cox regression models where it underestimates the treatment effect. The key point is that the first stage residual adds individual variability (a frailty) in the Cox model estimated in the second-stage.\par

\section{Monte Carlo Simulation Study}
\label{sec:montecarlo}
  
To evaluate the behavior of the proposed methodology in finite samples, we conducted a range of Monte Carlo simulations.
We found that, beyond the expected effect on precision of estimation,
neither the baseline shape of the survival times nor the censorship distribution have any meaningful effect on the observed results. Therefore, we only show results when the baseline survival times follow a Weibull distribution with shape parameter two and scale parameters coherent with the proportional hazard assumption; that is, the scale parameter equates to $\exp\{1+\beta_X\cdot (X-\bar X) + Z + \beta_U\cdot U\}$. Here $\beta_X$ is the target. We subtract $\bar X$ from $X$, as opposed to drawing $X$ from a distribution with mean 0, to ensure that we obtain realistic survival times with binary $X$ without needing to alter the intercept. $Z$ and $U$ denote the measured and unmeasured confounders, respectively. 
Both the measured ($Z$) and unmeasured ($U$) covariates follow independent standard normal distributions, $\mathscr N_{0,1}$. Censoring was independently drawn from a 
Weibull distribution such that the expected censorship was 20\%. Treatment assignment is based on the linear equation 
$X^*=1+W+Z+\alpha_V\cdot V + \epsilon$, where $W$ is the instrument, $V$ is the unmeasured covariate, and $\epsilon$ is the random noise. We set $X=X^*$ for a continuous exposure and $X=I(X^* \geq 0)$ for a binary exposure. All of $W$, $V$, and $\epsilon$ are drawn from independent standard-normal distributions. Notice that, after fixing the rest of the parameters, increasing $\alpha_V$ yields an instrumental variable of lesser quality.  Sample size was fixed at $N=500$.\par

\subsection{All Omitted Covariation is Unmeasured Confounding}

We first suppose $X=X^*$, $U=V$ and $\epsilon$ follow possibly correlated standard 
normal distributions. That is, the endogenous variable is continuous and there are no unmeasured predictors of survival time unrelated with treatment selection (i.e., while the true Cox model of the survival times may include a shared covariate with the treatment selection process, it does not include a frailty).

\begin{figure}[b!]
\begin{center}
\begin{tabular}{c}
\includegraphics[width=14cm]{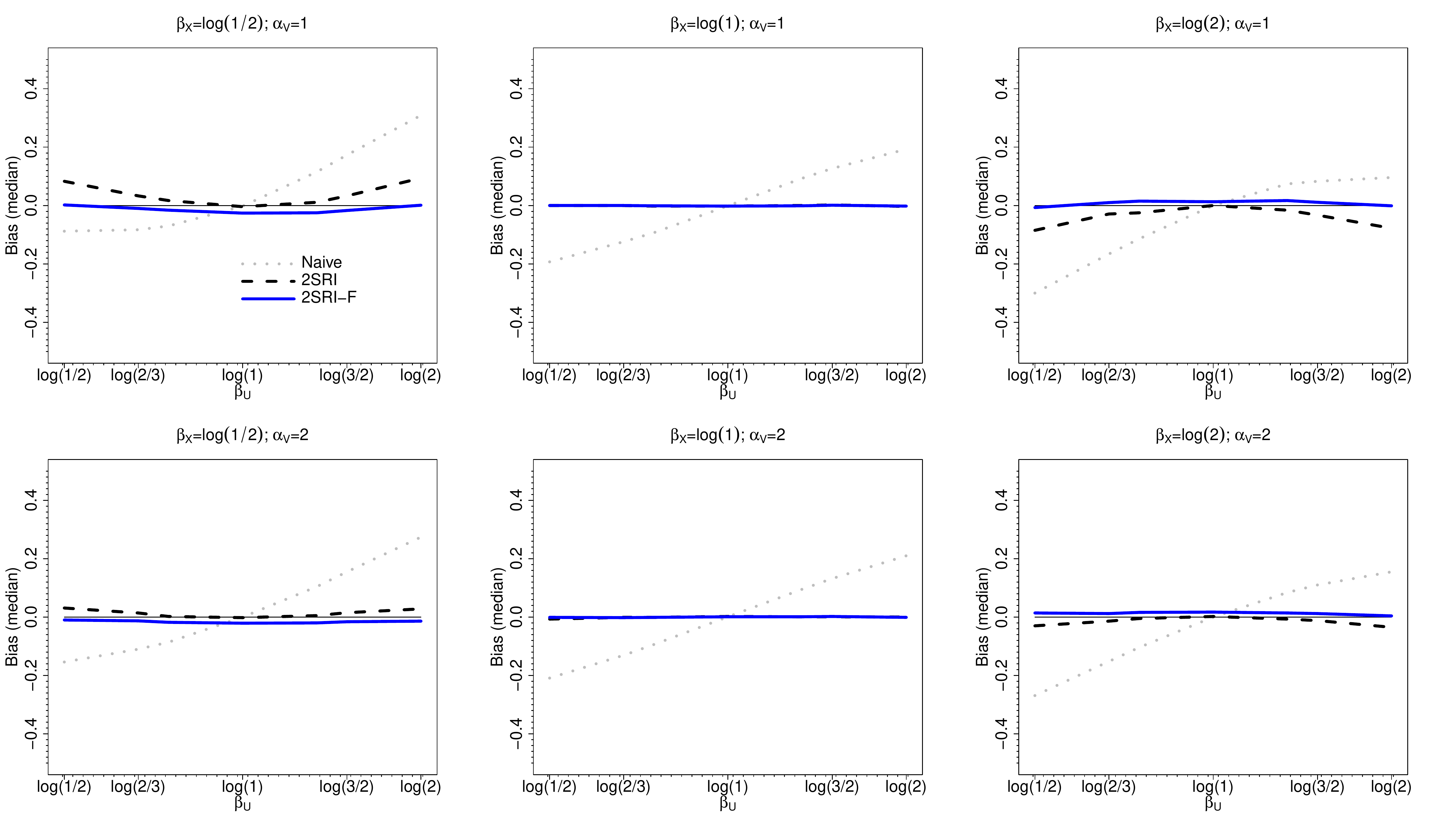}\\
\includegraphics[width=14cm]{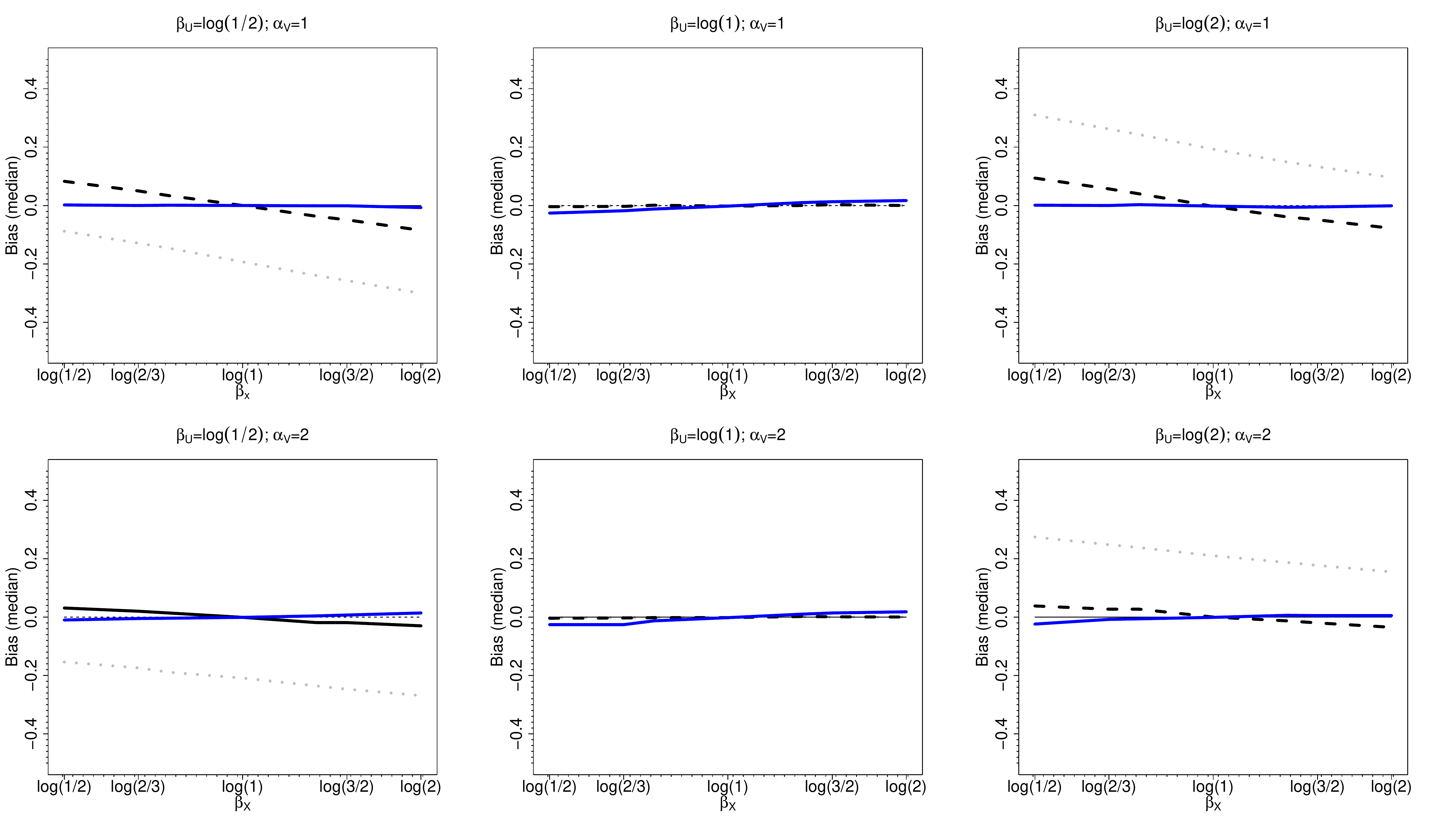}\\
\end{tabular}
\end{center}
\caption{Median bias for Weibull($\exp\{1+\beta_X\cdot X + Z + \beta_U\cdot U\}$, 2) times and 
$X=X^*=1+Z+W+\alpha_V\cdot U+\epsilon$, where $Z$, $U$, $W$ and $\epsilon$ follow independent standard normal distributions. 
Gray-dotted, naive Cox model (omitted covariate is ignored); black-dashed 2SRI, procedure; blue-continuous, 2SRI-F (2SRI plus Gaussian frailty). Continuous black thin line stands for the zero-bias situation.}
\label{scenarioI}
\end{figure}

Figure \ref{scenarioI} shows the median of the bias observed in 2000 Monte Carlo iterations for a stronger, $\alpha_V=1$, and weaker, $\alpha_V=2$, instrument. The Naive Cox model, which ignores the presence of omitted covariates, only performed well for $\beta_U=0$ (there are no omitted covariates and the Cox model is correct). The proposed algorithm, 2SRI-frailty (2SRI-F), reduced bias the most when the omitted covariates had strong effects. When the presence of unmeasured confounding was weaker ($\beta_U$ close to zero), and when there was no effect of 
the treatment, $\beta_X=0$, both 2SRI and 2SRI-F obtained similar results. Median-bias appeared to be invariant to the strength of the instrument.\par

\begin{center}
\begin{table}
\caption{Observed coverage of 95\% confidence intervals for the Cox model treatment effect covariate, Naive, 2SRI algorithm and the proposed 
2SRI algorithm with gaussian individual frailty (2SRI-F).\label{T1}}
\begin{center}
\begin{footnotesize}
\begin{tabular}{cccccccccccccccccccccccccccccccc}
\multicolumn{13}{l}{\rule{0.75\linewidth}{1pt}}\\                      
&             &                  && \multicolumn{3}{c}{$\alpha_{U_X}=1$} && \multicolumn{3}{c}{$\alpha_{U_X}=2$}\\ \cline{2-11}
&{$e^{\beta_U}$ } & {$e^{\beta_X}$}      && {\bf Naive} & {\bf 2SRI} & {\bf 2SRI-F} && {\bf Naive} & {\bf 2SRI} & {\bf 2SRI-F}\\  \cline{2-3}\cline{5-7} \cline{9-11}          
& $1/2$       &   $1/2$          && 0.466 & 0.671 & 0.871 && 0.019 & 0.867 & 0.891\\
&             &   $1$            && 0.000 & 0.931 & 0.931 && 0.000 & 0.917 & 0.922\\  
&             &   $2$            && 0.000 & 0.670 & 0.884 && 0.000 & 0.868 & 0.898\\ 
& $1$         &   $1/2$          && 0.952 & 0.951 & 0.912 && 0.952 & 0.851 & 0.912\\
&             &   $1$            && 0.940 & 0.948 & 0.949 && 0.940 & 0.948 & 0.949\\
&             &   $2$            && 0.943 & 0.947 & 0.915 && 0.943 & 0.947 & 0.915\\ 
& $2$         &   $1/2$          && 0.000 & 0.617 & 0.857 && 0.000 & 0.861 & 0.874\\
&             &   $1$            && 0.000 & 0.935 & 0.940 && 0.000 & 0.902 & 0.907\\
&             &   $2$            && 0.396 & 0.685 & 0.992 && 0.012 & 0.878 & 0.898\\ 
\multicolumn{13}{l}{\rule{0.75\linewidth}{1pt}}\\ 
\end{tabular}
\end{footnotesize}
\end{center}
\end{table}
\end{center}

Table \ref{T1} reports the coverage of the 95\% confidence intervals computed from the standard asymptotic variance obtained from the second-stage 
Cox regression models of the respective procedures, ignoring the first stage variability.The proposed algorithm achieved coverage close to the nominal level in all cases, suggesting that it will be able to be implemented easily in practice. In contrast, the coverage of the naive Cox methods and the basic 2SRI algorithm was poor. 
As expected, the naive method performed correctly for $\beta_U=0$, the case when the model is correct. As is well-known, 
two-stage instrumental variable methods lead to estimators ($\hat\beta^{IV}_X$) with the degree of variance inflation depending on the strength of the instrumental variable. We found that the length of the 95\% confidence intervals 
ranged between 0.09 and 0.17 for the naive models, between 0.20 and 0.23 for 2SRI and between 0.21 and 0.26 for the proposed 2SRI-F method, 
respectively, for both $\alpha_{V}=1$ and $\alpha_{V}=2$. The fact that the 2SRI-F procedure imposes a greater amount of inflation compared to 2SRI is helpful in terms of its ability to maintain the nominal level of coverage. Crucially, it appears that the 2SRI-F's assumption of the distribution of the frailty is helpful in addressing bias and does not suffer from the excessive and inappropriate gains in precision that accompany many procedures with parametric components.\par

\subsection{Omitted Covariation is a Mixture of Unmeasured Confounding and a Pure Individual Frailty}

\begin{figure}[b!]
\begin{center}
\includegraphics[width=10cm]{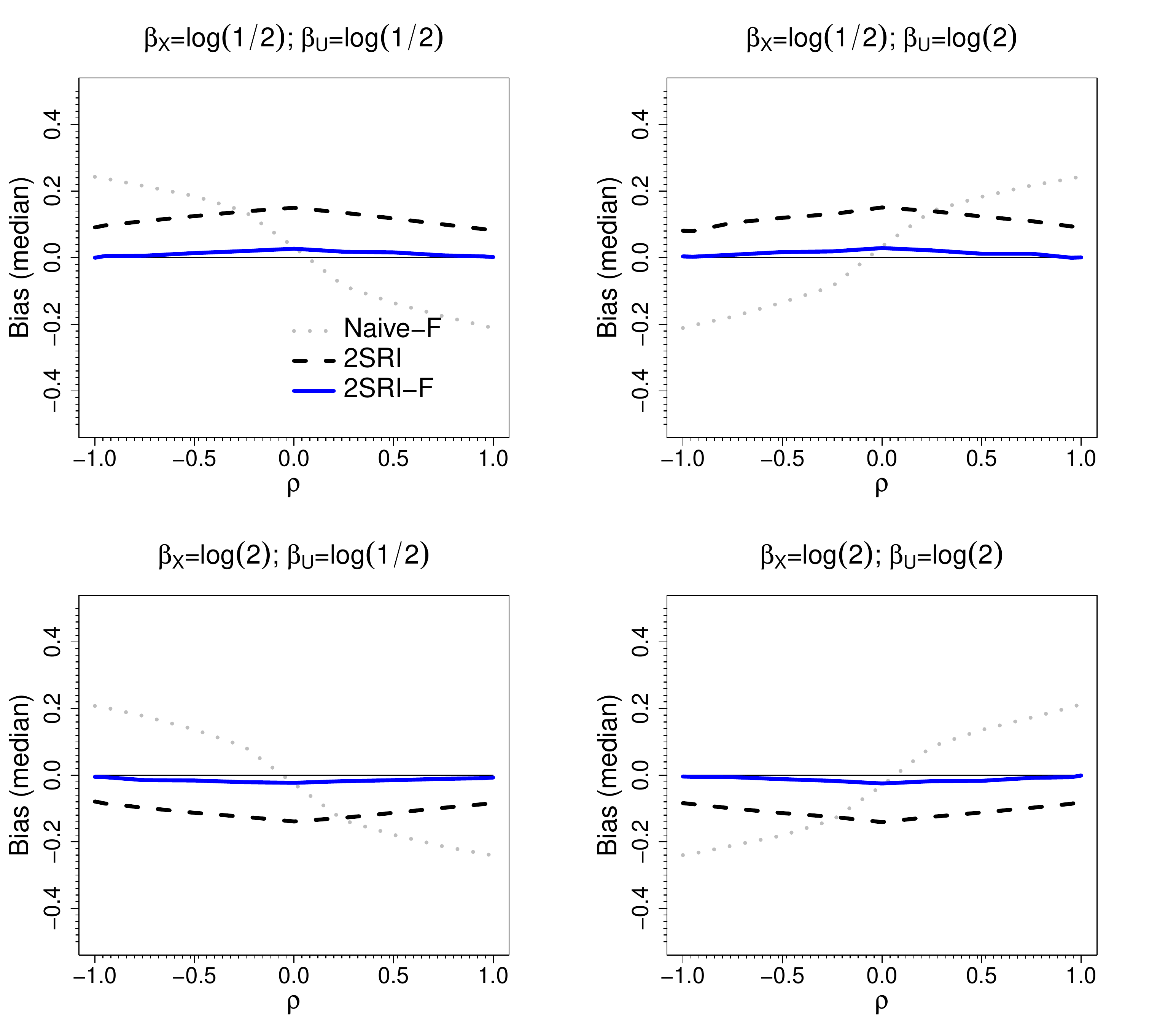}
\end{center}
\caption{Median bias for Weibull($\exp\{1+\beta_X\cdot X + Z + \beta_U\cdot U\}$, 2) times and 
$X=1+Z+W+U+\epsilon$, $Z$, $U$, $W$ and $\epsilon$ follow standard normal distributions. Correlation between $U$ and $V$ is $\rho$.
Gray-dotted, naive Cox model with a Gaussian frailty term (omitted covariate is ignored but a Gaussian frailty is included in the model); black-dashed 2SRI, procedure; blue, 2SRI-F (2SRI plus Gaussian frailty). Continuous black thin line stands for the zero-bias situation.}
\label{correlations}
\end{figure}

Figure \ref{correlations} depicts the observed median bias for the previous scenario for $\alpha_V=1$ and cor$(U,V)=\rho$. Note that $\rho=0$ implies 
that the survival model does not contain unmeasured confounders, only unmeasured covariates, while $\rho= \pm 1$ implies that all omitted covariation manifests as unmeasured confounding (i.e., is also related to treatment assignment). In the $\rho=0$ case, it is reassuring that the 2SRI and the 2SRI-F procedures perform nearly as well as the naive Cox regression model, the true model in this scenario. The advantage of using 2SRI-F versus 2SRI was larger for $\rho$ close to zero, which makes sense as the pure frailty variation is at its maximum, whereas at values close to $\pm 1$ the frailty has all but disappeared.

In order to check the robustness of the recommended gaussian frailty with  respect to the unmeasured covariates distribution, we study the case where the unmeasured covariates, $U$ and $V$, are not normally distributed. In particular, we considered the following scenarios:\par
\medskip 
{\bf M-I.}\,\,\,\, $U=\sqrt{a}\cdot\gamma_1 + \sqrt{(1-a)}\cdot\gamma_2;\, V=\sqrt{a}\cdot\gamma_1 + \sqrt{(1-a)}\cdot\gamma_3$.\par
{\bf M-II.} \,   $U=\sqrt{a}\cdot\gamma_1 + \sqrt{(1-a)}\cdot \eta_1;\, V=\sqrt{a}\cdot\gamma_1 + \sqrt{(1-a)}\cdot \eta_2$.\par
{\bf M-III.}     $U=\sqrt{a}\cdot \eta_1 + \sqrt{(1-a)}\cdot \eta_2;\, V=\sqrt{a}\cdot \eta_1 + \sqrt{(1-a)}\cdot \eta_3$.\par
{\bf M-IV.}\,    $U=\sqrt{a}\cdot \eta_1 + \sqrt{(1-a)}\cdot \gamma_2;\, V=\sqrt{a}\cdot \eta_1 + \sqrt{(1-a)}\cdot \gamma_3$.\par
\begin{figure}[b!]
\begin{center}
\includegraphics[width=10cm]{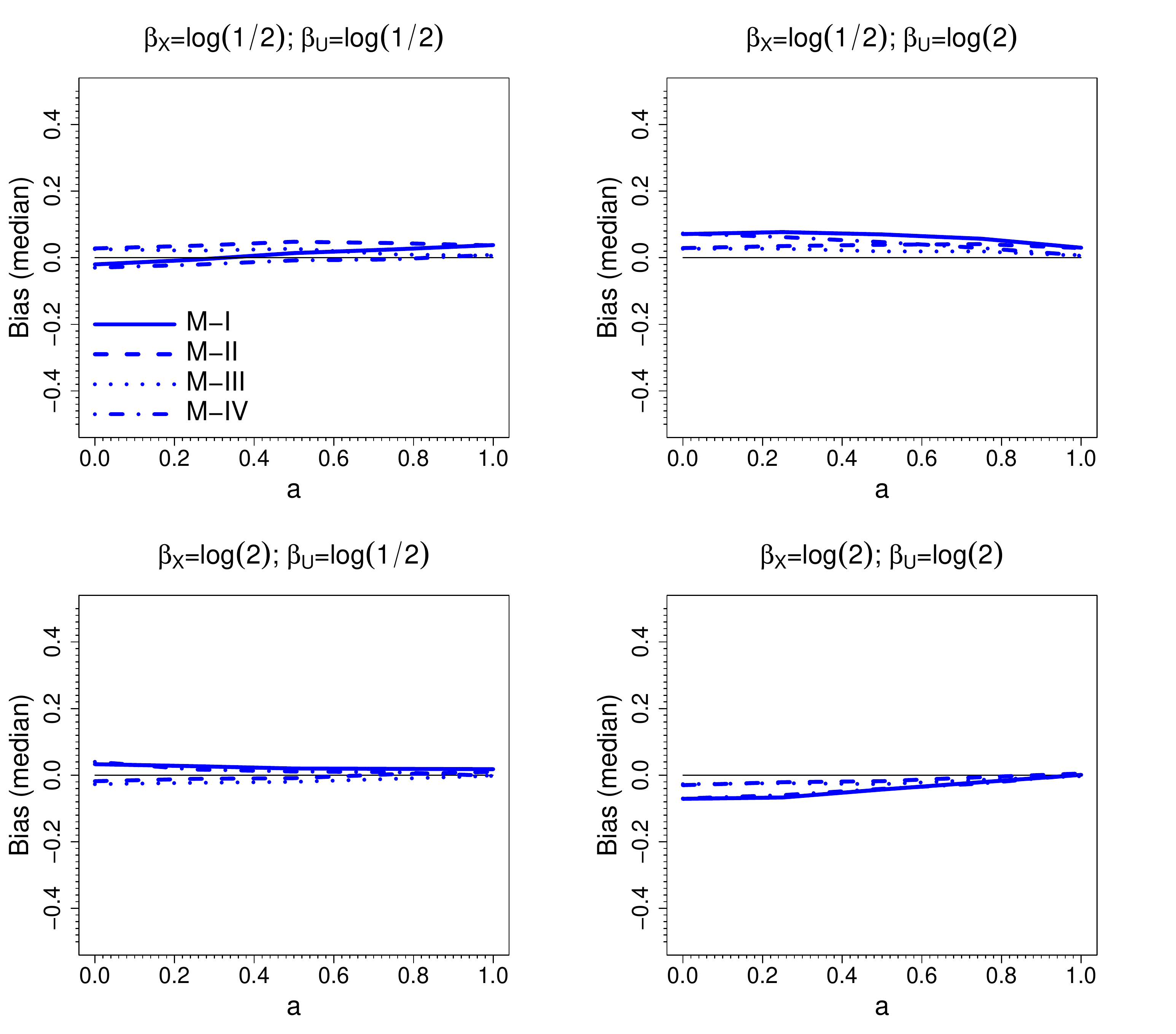}
\end{center}
\caption{Median bias for Weibull($\exp\{1+\beta_X\cdot X + Z + \beta_U\cdot U\}$, 2) times and 
$X=1+Z+W+U+\epsilon$, $Z$ and $\epsilon$ follow standard normal distributions and $U$ and $V$ following previous models always 
using the 2SRI-F algorithm. Continuous black thin line stands for the zero-bias situation.}
\label{fap}
\end{figure}

\medskip
\noindent
where $\gamma_1$, $\gamma_2$ and $\gamma_3$ follow independent centered Gamma(1,1) and $\eta_1$, $\eta_2$ and $\eta_3$ follow independent 
standard normal distribution. Note that parameter $a$ determines both the distribution and the relationship between $U$ and $V$ and it is chosen in order to keep constant the marginal variance. Figure \ref{fap} shows the median of the bias of 2SRI-F algorithm observed in 2000 Monte Carlo iterations under 
the previous models with $a\in [0,1]$. Results suggest minimal impact of the unmeasured covariates distribution. As expected, when the frailty has the assumed distribution the bias is smaller but, crucially, observed biases were always smaller than under the 2SRI procedure.\par

\subsection{Nonlinear Treatment Selection Model: Binary Exposure}

The second scenario supposes $X=I_{[0,\infty)}(X^*)$, a binary exposure. Because other values of  $\rho$ produced similar results, we only report results for which the correlation between $U$ and $V$ was fixed at $\rho=1/2$.

\begin{figure}
\begin{center}
\begin{tabular}{c}
\includegraphics[width=14cm]{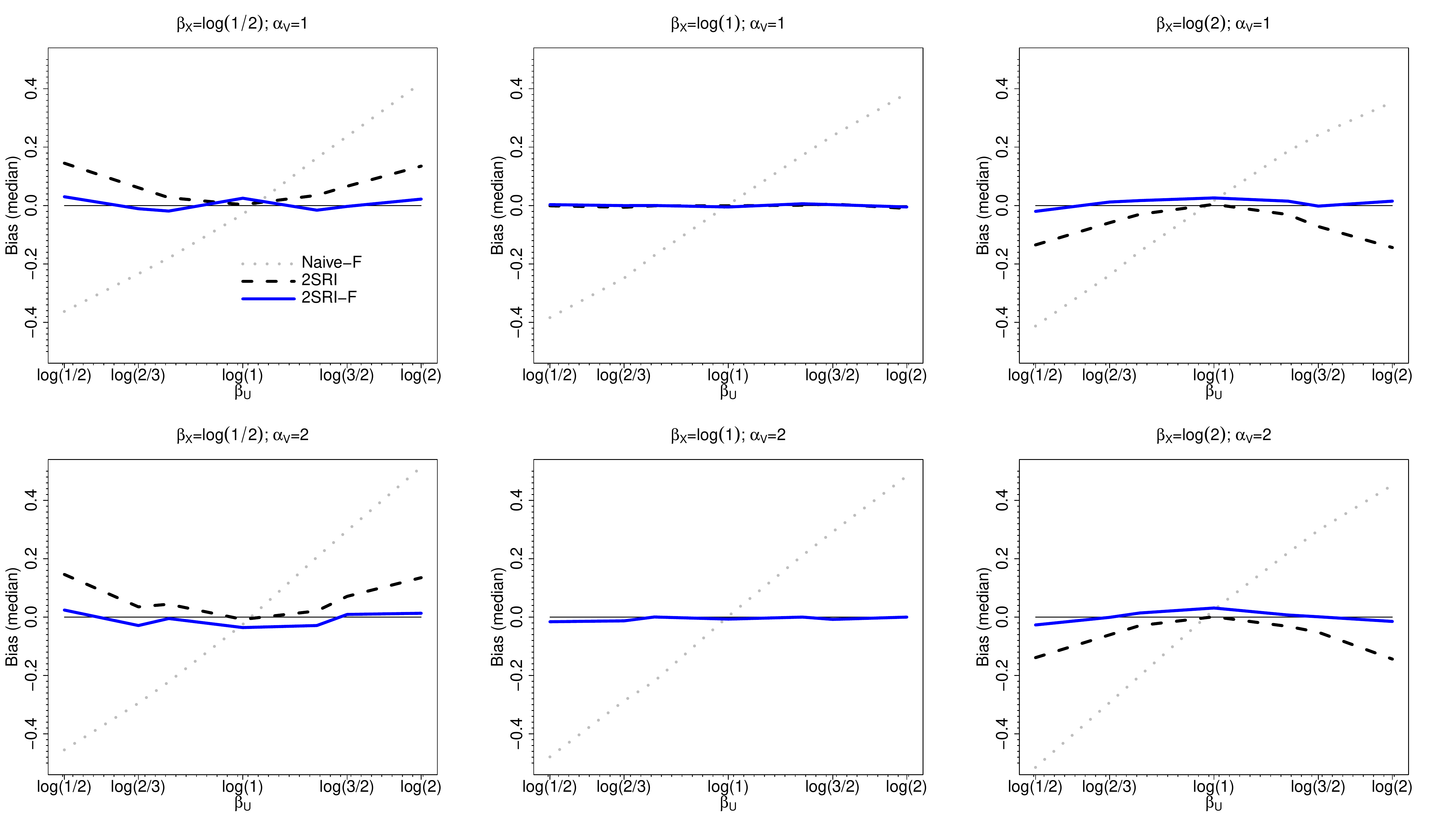}\\
\includegraphics[width=14cm]{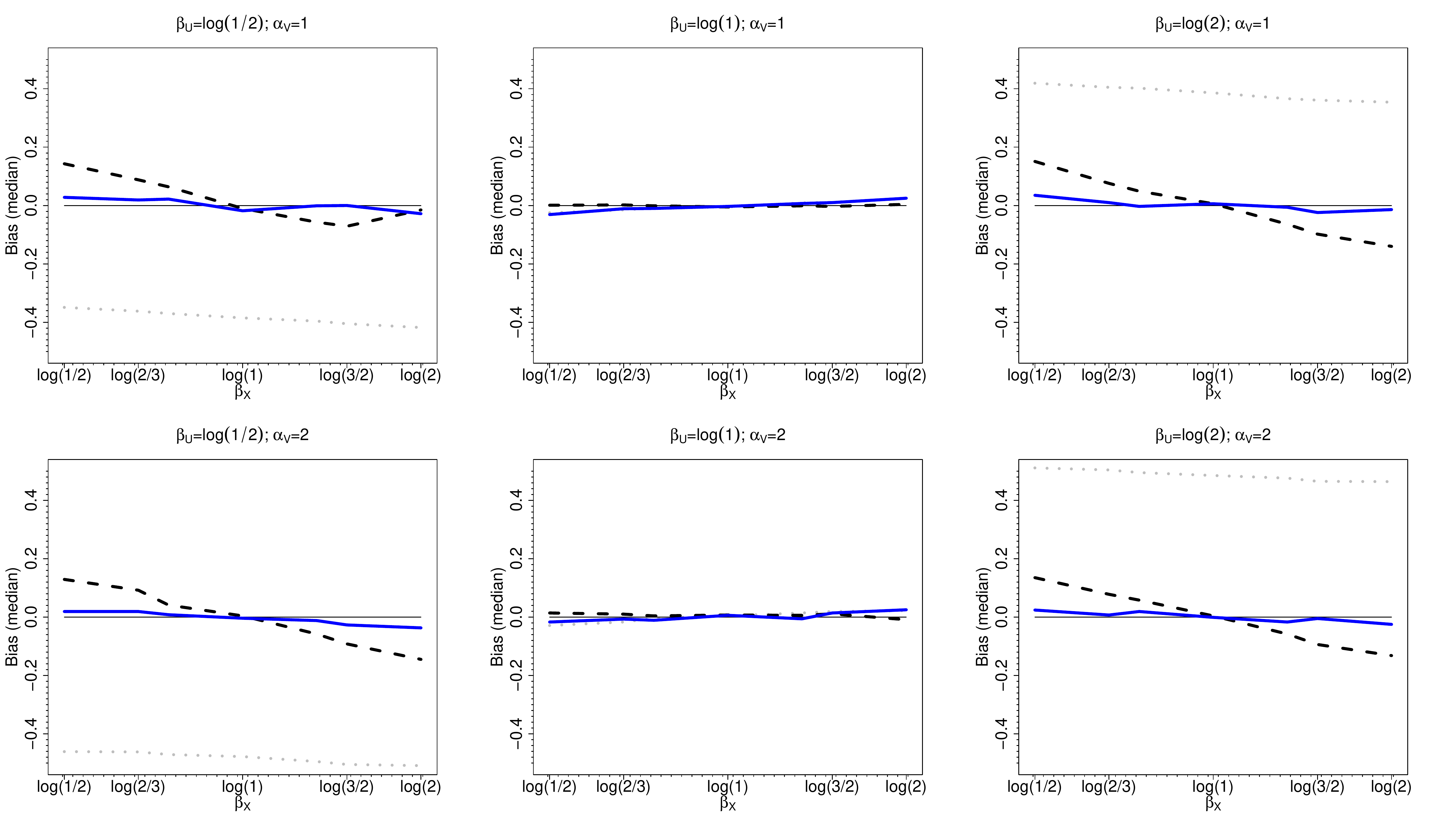}\\
\end{tabular}
\end{center}
\caption{Median bias assuming Weibull($\exp\{1+\beta_X\cdot X + Z + \beta_U\cdot U\}$, 2) survival times and 
$X^*=1+Z+W+\alpha_V\cdot U+\epsilon$, with $X=I_{[1,\infty)}(X^*)$, $Z$, $U$, $W$ and $\epsilon$ following standard normal distributions. Correlation between $U$ and $V$ is 1/2.
Legend: Gray-dotted, naive Cox model with Gaussian frailty (omitted covariate is ignored but a Gaussian frailty is included); black-dashed 2SRI, procedure; blue, 2SRI-F (2SRI plus Gaussian frailty). Continuous black thin line stands for the zero-bias situation.}
\label{scenarioII}
\end{figure}

Figure \ref{scenarioII} depicts the median bias over 2000 Monte Carlo iterations when $\beta_X$ was directly estimated from Cox regression with Gaussian frailty, (due to the presence of an unmeasured covariate
unrelated with treatment assignment, this model is the correct model in the absence of unmeasured confounding), the 2SRI procedure, and 2SRI-F. A stronger ($\alpha_V=1$) and a weaker ($\alpha_V=2$) scenario was considered for the instrument. Not surprisingly, ignoring the presence of the frailty and estimating a standard Cox regression model results in larger bias, even in the absence of an unmeasured confounder. 
The 2SRI algorithm helps us to control just the part of the bias related with the treatment assignment but also fails to handle the frailty and its performance suffers as a result. The naive Cox model with a frailty performs much better than both the naive Cox model with no frailty and 2SRI, implying that accounting for the frailty may be more important than dealing with unmeasured confounding. However, the proposed 2SRI-F methodology produces a yet further reduction in bias, to close to zero in all scenarios, which we conjecture is due to separating the idiosyncratic and confounding effects of $U$. These results reveal that there is clear benefit to be gained in practice from using 2SRI-F as an IV procedure for time-to-event data.\par
\medskip

\begin{center}
\begin{table}
\caption{Observed coverage of 95\% confidence intervals for the Cox model treatment effect covariate, Naive, 2SRI algorithm and the proposed 
2SRI algorithm with gaussian individual frailty (2SRI-F).\label{T2}}
\begin{footnotesize}
\begin{center}
\begin{tabular}{cccccccccccccccccccccccccccccccc}
\multicolumn{13}{l}{\rule{0.8\linewidth}{1pt}}\\                      
&             &                  && \multicolumn{3}{c}{$\alpha_{U_X}=1$} && \multicolumn{3}{c}{$\alpha_{U_X}=2$}\\ \cline{2-11}
&{$e^{\beta_U}$ } & {$e^{\beta_X}$}      && {\bf Naive-F} & {\bf 2RSI} & {\bf 2RSI-F} && {\bf Naive-F} & {\bf 2RSI} & {\bf 2RSI-F}\\  \cline{2-3}\cline{5-7} \cline{9-11}          
& $1/2$       &   $1/2$          && 0.325 & 0.885 & 0.943 && 0.102 & 0.916 & 0.947\\
&             &   $1$                  && 0.189 & 0.936 & 0.944 && 0.041 & 0.944 & 0.947\\
&             &   $2$                  && 0.158 & 0.885 & 0.938 && 0.037 & 0.925 & 0.946\\ 
& $1$         &   $1/2$           && 0.941 & 0.954 & 0.949 && 0.936 & 0.944 & 0.943\\
&             &   $1$                 && 0.946 & 0.943 & 0.947 && 0.950 & 0.949 & 0.947\\
&             &   $2$                 && 0.936 & 0.946 & 0.944 && 0.939 & 0.905 & 0.947\\ 
& $2$         &   $1/2$           && 0.158 & 0.883 & 0.936 && 0.036 & 0.943 & 0.942\\
&             &   $1$                 && 0.212 & 0.940 & 0.946 && 0.038 & 0.927 & 0.942\\
&             &   $2$                 && 0.321 & 0.880 & 0.933 && 0.099 & 0.914 & 0.948\\ 
\multicolumn{13}{l}{\rule{0.8\linewidth}{1pt}}\\ 
\end{tabular}
\end{center}
\end{footnotesize}
\end{table}
\end{center}

Both the 2SRI and the proposed algorithms achieved coverage close to the nominal level in all scenarios; the naive Cox model with Gaussian frailty (Naive-F) understandably yields good results only when $\beta_U=0$ (Table \ref{T2}). As expected, the strength of the instrument affects the confidence interval widths. The width of the 95\% confidence interval ranged between 0.44 and 0.57 for the Naive-F models. Under $\alpha_{V}=1$ interval estimatior width ranged between 0.86 and 0.88 and between 0.91 and 1.07 for the 
2SRI and 2SRI-F, respectively. For $\alpha_{V}=2$ the widths ranged between 1.23 and 1.31 
and between 1.30 and 1.53 for 2SRI and 2SRI-F, respectively. These results are consistent with the results for continuous exposures; the variance inflation under the 2SRI-F procedure exceeds that under 2SRI which in-turn exceeds that under Naive-F.

\section{Real-world application: The Vascular Quality Initiative dataset}
\label{sec:application}

We apply 2SRI-frailty to nationwide data from the Vascular Quality Initiative (VQI) (\url{http://www.vascularqualityinitiative.org}) on patients diagnosed with carotid artery disease (carotid stenosis). These data contain comprehensive information on all patients suffering from carotid stenosis and is continually updated over time to facilitate determination of the best procedure or treatment approach to use on average and to determine which type of patients benefit the most from each procedure. However, the data are exposed to a plethora of selection biases raising concerns that naive analyses will yield biased results. Because the outcomes of most interest are events such as stroke or death that can occur at any point during follow-up, these data are ideal for application of the 2SRI-frailty procedure.

We employed 2SRI-F to estimate the comparative effectiveness of carotid endarterectomy (CEA) versus carotid angioplasty 
and stenting (CAS), the two surgical procedures used to intervene on patients with carotid stenosis. The data consist of 28712 patients who received CEA and 8117 who received CAS, between 15 and 89 years of age, over 2003-2015. During follow-up, there were 3955 and 807 deaths in the CEA and CAS groups, respectively. Table \ref{T3} shows descriptive statistics for the measured covariates by procedure.\par

\begin{center}
\begin{table}
\caption{Descriptive statistics for measured covariates both overall and by carotid endarterectomy (CES) versus carotid angioplasty 
and stenting (CAS) recipients.\label{T3}}
\begin{footnotesize}
\begin{center}
\begin{tabular}{clcccccccccccccccccccccc}
\multicolumn{13}{l}{\rule{0.7\linewidth}{1pt}}\\ 
&                                                        &&     {\bf CEA}     &&       {\bf CAS}  \\
&                                                        &&      n=28,712    &&       n=8,117 \\  \cline{2-2} \cline{4-6}
& {\bf Age, mean$\pm$sd}               &&  70.2$\pm$9.4 && 69.1$\pm$10.3\\
& {\bf Male, n (\%)}                         &&  17,180 (59.8)  && 5,119  (63.1)	\\
& {\bf Race, n (\%)}\\				
& {\phantom {bla}\bf White}            &&	27,033 (94.2) && 7,382 (90.9)\\
&  {\phantom {bla}\bf Black}            &&	922 (3.2)	  && 433 (5.3)\\
&  {\phantom {bla}\bf Other}          &&	757 (2.6)	 && 302 (3.8)\\
& {\bf Elective, n (\%)}                                 &&	24,906 (86.7)&& 6,587 (81.1)\\
& {\bf Symptomatic, n (\%)}                         && 	11,168 (38.9)&&4,282 (52,7)\\
& {\bf TIA or amaurosis, n (\%)}                  && 	6,405 (22.3)	&& 2,001 (24.6)\\
& {\bf Stroke, n (\%)}                                   &&	4,763 (16.6) && 2,281 (28.1)\\
& {\bf Hypertension, n (\%)}		    && 25,452 (88.6)&& 7,235 (89.1)\\
& {\bf Smoking History, n (\%)}		    && 22,098 (77.0)&& 6,168 (76.0)\\
& {\bf Positive Stress Test, n (\%)}             && 2,655 (9.2)     && 677 (8.3)\\
& {\bf Coronary Disease, n (\%)} 		   && 8,586 (29.9)	&& 2,790 (34.4)\\			
& {\bf Heart Failure, n (\%)}                      && 2,669 (9.3)      && 1,215 (15.0)\\
& {\bf Diabetes, n (\%)}                            && 9,749 (33.9)     && 2,942 (36.2)\\
& {\bf COPD, n (\%)}                                  && 6,229 (21.7)	&& 2,083 (25.7)\\
& {\bf Renal Insufficiency, n (\%)}           && 1,649 (5.7)	&& 446 (5.5) \\
& {\bf HD, n (\%)}                     && 	263 (0.9)       && 114 (1.4)\\
& {\bf Prior ipsilateral CEA, n (\%)}  && 4,472 (15.6)     && 2,857 (35.2)\\
& {\bf Antiplatelet therapy, n (\%)}\\			
& {\phantom {bla}\bf Aspirin}          && 23,960 (83.4)  && 6,932 (85.4)\\
& {\phantom {bla}\bf P2y12 inhibitor}  && 6,980 (24.3) &&	6,173 (76.0)\\
& {\bf Beta-blocker, n (\%)}           && 18,269 (63.6) && 4,602 (56.7)\\ 
& {\bf Statin, n (\%)}                             && 22,418 (78.1) && 6,408 (78.9)\\
\multicolumn{13}{l}{\rule{0.7\linewidth}{1pt}}\\ 
\end{tabular}
\end{center}
\end{footnotesize}
\end{table}
\end{center}

\begin{figure}[b!]
\begin{center}
\includegraphics[width=14cm]{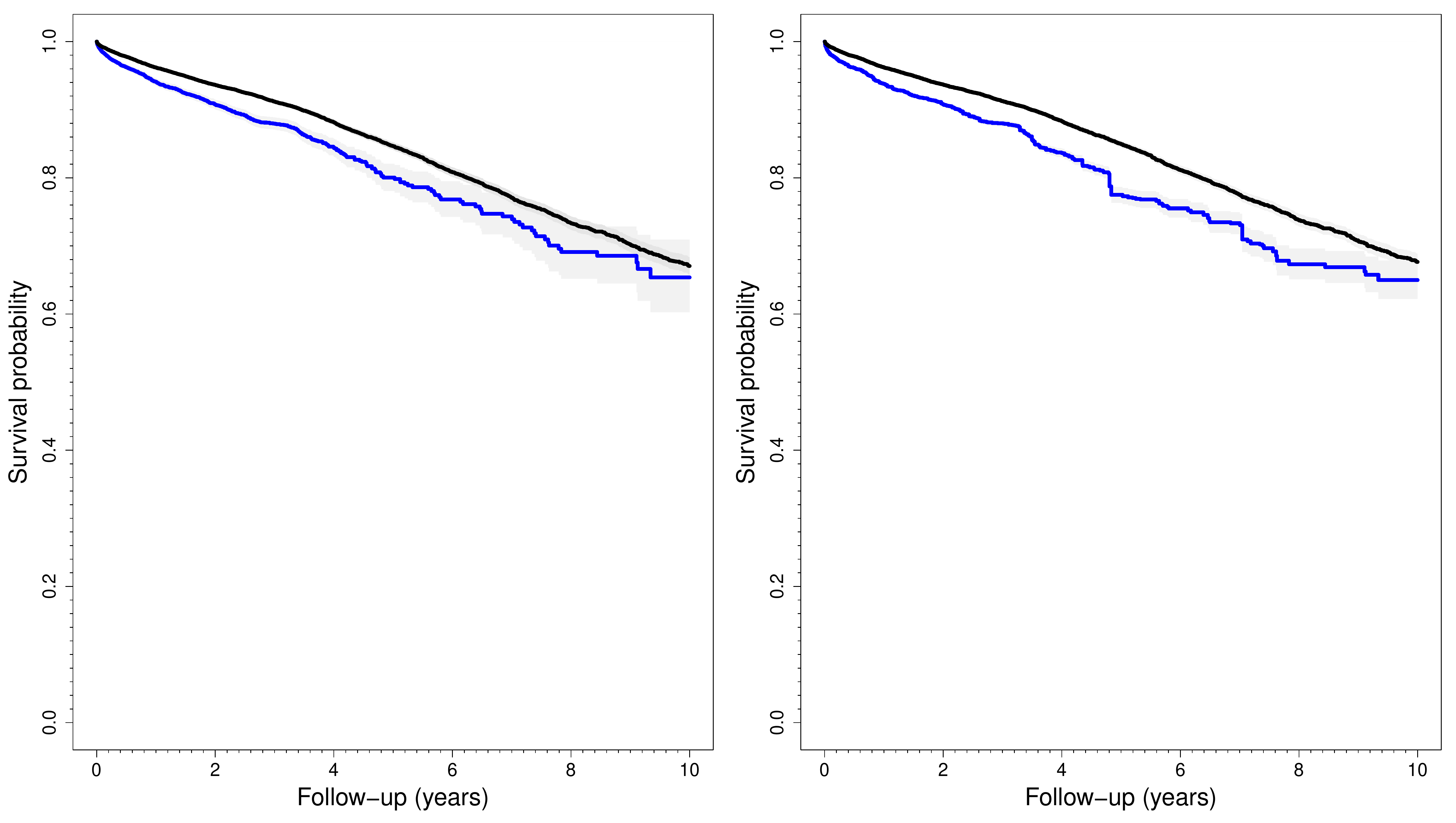}
\end{center}
\caption{Kaplan-Meier estimations for both CEA and CAS group with 95\% confidence bands: crude (left) and adjusting by the covariates described in table  \ref{T3} using the weighted inverse propensity procedure (right).}
\label{survivalCurves}
\end{figure}

Figure \ref{survivalCurves} depicts two Kaplan-Meier survival curves: crude (left) and adjusted by patient age, gender,  ethnicity, race. type of surgery (elective/not elective), symptons (yes/no),  hypertension diabetes, smoking history (yes/no), positive stress test, coronary 
disease, heart failure, diabetes, COPD, renal insufficiency, dialysis status (HD), prior ipsolateral CEA, and the use of antiplatelet therapy, beta-blokers and statin by using the weighted inverse propensity procedure \cite{todd12}. 
The crude hazard ratio (HR) comparing CEA to CAS was 0.719 (95\% CI of (0.666; 0.777)). Adjusted HR was 0.693 (0.633; 0.760). The last HR is slightly modified when a frailty term is included: 0.685 (0.624; 0.753), and 0.676 (0.613; 0.745) for the Gaussian and Gamma cases, respectively.\par

As the instrumental variable we used the center level frequency of CEA versus CAS procedures over the twelve months prior to the current patient; that is, total CEA divided by total of CEA and CAS procedures in the twelve months prior to the current patient. This variable is justified as an instrument due to: 1) hospitals that performed a high relative amount of a certain procedure in the past are likely to keep doing so; 2) there should be no effect of the relative frequency of CEA vs CAS on a patient's outcome except through its effect on treatment choice for that patient; 3) we know of no factors that would influence both this frequency and a patient's outcome. Reasons 2) and 3) are contingent on adjusting for the total number of CEA and CAS procedures performed at the center over the past 12 months.\par

\begin{figure}[b!]
\begin{center}
\includegraphics[width=14cm]{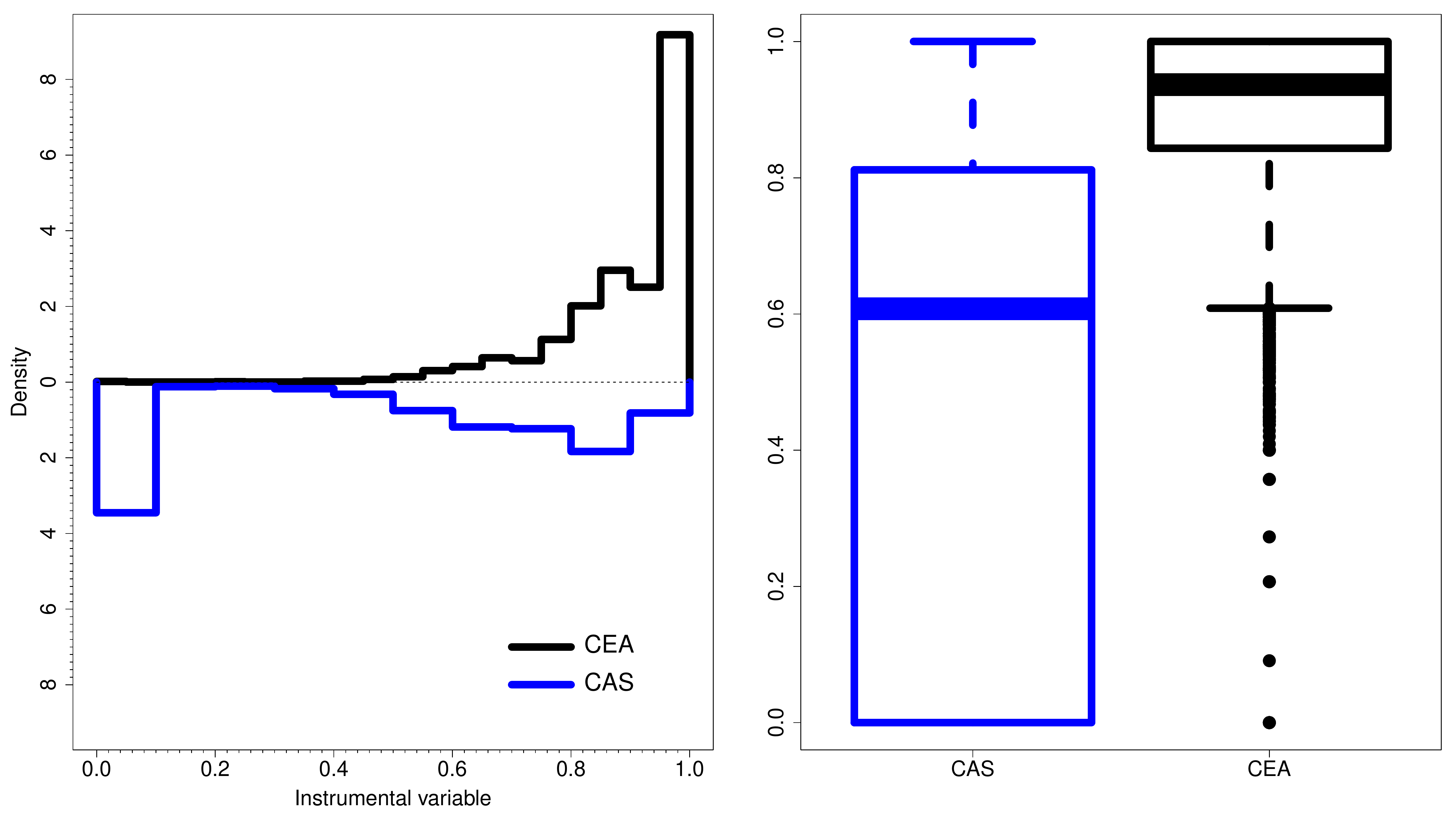}
\end{center}
\caption{Histograms for the instrument variable (left), $W$, and boxplot (right), by received treatment.}
\label{viDistribution}
\end{figure}

On the VQI data the IV is highly associated with treatment choice. The probability that a randomly selected subject undergoing CEA has 
a larger value of the instrument than a randomly selected subject undergoing CAS, was 0.881 (95\% confidence interval of (0.876; 0.885)). This IV was unrelated with all of the measured confounders suggesting anecdotally that it may also be uncorrelated with any unmeasured confounders. Hence, it is reasonable to assume that the relationship of the instrument with mortality is solely due to its relationship with the treatment. Figure \ref{viDistribution} (left side) shows the histogram of $W$ in both CEA and CAS groups, at right, we show the boxplot for the IV by surgical procedure.\par

\begin{table}
\begin{center}
\caption{Hazard ratios and 95\% confidence intervals by estimation method.\label{T4}}
\begin{footnotesize}
\begin{tabular}{clcccc}
\multicolumn{6}{c}{\rule{0.56\linewidth}{1pt}}\\                      
&                                                           && {\bf HR (95\% CI)}\\ \cline{3-4}
& {\bf Crude}                              && 0.719 (0.666; 0.777)\\
& {\bf Cox model (Naive)}                  && 0.693 (0.633; 0.760)\\
& {\bf Naive - frailty (gaussian)}         && 0.685 (0.624; 0.753)\\
& {\bf Naive - frailty (gamma)}            && 0.676 (0.613: 0.745)\\
& {\bf 2SRI}                               && 0.901 (0.737; 1.100)\\  
& {\bf 2SRI - frailty (gaussian)}          && 0.887 (0.724: 1.087)\\ 
& {\bf 2SRI - frailty (gamma)}             && 0.882 (0.716; 1.086)\\ 
\multicolumn{6}{c}{\rule{0.56\linewidth}{1pt}}\\ 
\end{tabular}
\end{footnotesize}
\end{center}
\end{table}

The treatment effect almost disappears when 2SRI is applied on the dataset: HR of 0.901 with a 95\% confidence interval (0.737; 1.100). When the proposed 2SRI-frailty algorithm is used a similar result obtains: a HR of 0.887 ($\hat\beta_X^{\text IV}=-0.120$) with a 95\% confidence interval (0.724; 1.087). Similar results were also obtained under a gamma distributed frailty instead of the gaussian frailty. Table \ref{T4} shows the hazard ratios and 95\% confidence intervals.\par

\section{Discussion}
\label{sec:discussion}
Instrumental variables methods are often used to account for unmeasured confounders. Although these methods have 
been widely studied in a variety of situations, their suitability for estimating Cox proportional hazard models is unclear. It is well-known that, in this case, model misspecification 
can produce bias even when the omitted variables are unrelated with the treatment assignment \citep{R14}; that is, when they only affect the survival time. As suggested by our structural 
argument in the Introduction, an individual frailty appears able to solve this problem. We showed that the presence of idiosyncratic variation affecting treatment selection may induce a 
frailty in the instrumented survival time model even if there is no frailty in the original survival model. In practice, the most likely scenario is that both a true frailty and unmeasured 
confounding factors affect survival. For these reasons, we were motivated to develop and evaluate an IV procedure that, in the second stage, incorporates a frailty term.
 
Because the Cox model is nonlinear, our base strategy for dealing with unmeasured confounders was to use the two-stage residual inclusion algorithm, 2SRI, adapted to the Cox model. As noted above, even when the true survival model does not contain omitted covariates, the 2SRI procedure induces a frailty in the second-stage Cox regression model from the inclusion of the residuals computed in the first stage. To account for this phenomenon, we added an individual frailty in the second-stage (instrumented) statistical model. Under standard reliability conditions, we proved the asymptotic consistency 
of the estimator defined under our 2SRI-F procedure for the case when the univariate frailty distribution is correctly assumed to be Gaussian. 

Monte Carlo simulations suggested that the proposed methodology (2SRI-F) produces an 
important bias reduction and is superior to the 2SRI, particularly in the presence of an individual frailty due to unmeasured covariates unrelated with the treatment assignment. A very important finding is that the bias of the 2SRI-F method was always close to zero even when the residuals from the treatment selection equation were not normally distributed. The Gaussian distribution can be directly justified when each individual frailty is the sum of different independent sources of heterogeneity.  Furthermore, because the procedure with the Gaussian frailty was surprisingly robust to erroneously assumed frailty distributions, we recommend using a Gaussian frailty.

A controversial feature of our procedure is the inclusion of the individual frailty term. Although there exists a vast literature for the case where the frailty is 
common to a group of individuals (shared frailty), the number of references dealing with individual frailties is minimal. Consistency properties of the common estimation 
algorithm for Cox models with frailties were proved previously \citep{R22, R23}. We adapted these theoretical results in deriving the consistency results presented herein. By specifying a distribution for its values, the individual frailty accounts for the omitted covariates unrelated with the treatment assignment -- the extra variability introduced in the survival model from the first stage of the algorithm ($\epsilon$) and the {\it portion} of $V$ independent of $U$, freeing the augmented first-stage residual (the control function) to deal with unmeasured confounding. The resulting procedure estimates the 
average treatment effect conditional on the unmeasured confounder and the frailty \citep{R24}. Because in practice specification of the distribution of the frailty can be arbitrary, the observed results should be handled with caution \citep{R25} and they should be supported by sensitivity analyses considering different frailty 
distributions. In the VQI application, the empirical results were found to only have a slight dependence on the 
distribution of the frailty (see Figure \ref{fap}). This is a key finding that justifies the use of the 2SRI-F procedure and represents a major advance in the instrumental variables literature for the analysis of time-to-event outcomes in observational settings.\par

In the real-world application, a small but significant (at the 0.05 level) effect of the treatment is detected when the presence 
of omitted covariates on the Cox model is ignored. This effect almost disappears under 2SRI. When the 2SRI-F method is used, the estimated effect of CEA over CAS is slightly larger. This result confirms that the effect of the procedure a patient receives is underestimated when unmeasured confounding is ignored \citep{R26, R27}. It is
worth noting that different patient enrollment rates were observed by treatment: while CAS patient censorship is constant across the follow-up, most of the CEA patients have follow-up above two years with an important number censored between the second and the fourth years. To the extent these differences are caused by an unmeasured confounder, 
this can introduce additional bias in the standard naive estimates and strongly motivates the use of an adequate instrumental variable procedure.\par 

The method we developed conditions on all omitted covariates and assumes they have multiplicative effects on the hazard function under the Cox model, unlike recently developed methods 
that make unusual additive hazard assumptions in order to more simply account for unmeasured confounding \citep{todd1, R12}. Therefore, we anticipate that our proposed and proven procedure will hold extensive appeal and be widely used in practice.\par

While it is encouraging that our null results cohere with those of recent RCTs \citep{nejm, otru}, thereby overcoming the unfavorable CAS results of the non-IV analyses, an effect close to 0 makes it difficult to distinguish our proposed IV procedure from the incumbent two-stage residual inclusion method. However, when the true effect is 0 (HR of 1), the bias from ignoring the frailty is 0 due to the fact that omitting a frailty shrinks the true coefficient towards 0. Therefore, the differences between the 2SRI-F and standard 2SRI procedure for the Cox model are expected to converge to 0 as the true treatment effect approaches 0. In this sense, the lack of extensive differences between the various 2SRI (frailty and standard) procedures is a real-data endorsement that our proposed 2SRI-F procedure for the Cox model performs as it should by not rejecting the null hypothesis when the RCT results and the 2SRI results suggest that the true effect is close to 0.

\section*{Acknowledgement}
 This work was supported by a Patient-Centered Outcomes Research Institute (PCORI) Award ME-1503-28261. 
 All statements in this paper, including its findings and conclusions, are solely those of the authors and do 
not necessarily represent the views of the Patient-Centered Outcomes Research Institute (PCORI), its Board of Governors or Methodology Committee. 
The authors want to show their most sincerely grateful to the PCORI Patient Engagement and Governance Committee and specially to Jon Skinner for reading a draft of the manuscriptfor their efforts in the manuscripts revision and their help in developing the research proposal. The authors have no conflicts of interest to report.

\bibliographystyle{Chicago}
\bibliography{Bibliography}
\end{document}